\documentclass[3p,times,onecolumn,number]{elsarticle}
\usepackage{amsmath}
\usepackage[dvipdfm]{hyperref}

\newcommand{\bear}{\begin{eqnarray}}
\newcommand{\eear}{\end{eqnarray}}
\newcommand{\be}{\begin{equation}}
\newcommand{\ee}{\end{equation}}
\newcommand{\beqn}{\begin{eqnarray}}
\newcommand{\eeqn}{\end{eqnarray}}
\newcommand{\beqnn}{\begin{eqnarray*}}
\newcommand{\eeqnn}{\end{eqnarray*}}

\journal{...}

\begin{document}

\begin{frontmatter}
\title{Controlling Statistical Properties of a Cooper Pair Box Interacting with a Nanomechanical Resonator}

\author[al,CV,BB]{C.~Valverde\corref{cor1}}
\ead{valverde@unip.br}

\author[BB]{A.T.~Avelar},

\author[BB]{B.~Baseia}

\cortext[cor1]{Corresponding author}

\address[al]{Universidade Paulista, Rod. BR 153, km 7, 74845-090 Goi\^ania, GO, Brazil.}

\address[CV]{Universidade Estadual de Goi\'as, Rod. BR 153, 3105, 75132-903 An\'apolis, GO, Brazil.}

\address[BB]{Instituto de F\'{\i}sica, Universidade Federal de Goi\'as, 74001-970 Goi\^ania, GO, Brazil.}

\begin{abstract}

We investigate the quantum entropy, its power spectrum, and the excitation
inversion of a Cooper pair box interacting with a nanomechanical resonator, the first
initially prepared in its excited state, the second prepared in a
``cat"-state. The method uses the Jaynes-Cummings model with damping, with
different decay rates of the Cooper pair box and distinct detuning
conditions, including time dependent detunings. Concerning the entropy, it
is found that the time dependent detuning turns the entanglement more stable
in comparison with previous results in literature. With respect to the
Cooper pair box excitation inversion, while the presence of detuning
destroys the its collapses and revivals, it is shown that with a convenient
time dependent detuning one recovers such events in a nice way.

\end{abstract}

\end{frontmatter}

\section{Introduction}

In the last years there has been a great interest in the production of new
nonclassical states of the quantized electromagnetic field, one of the
interesting topics of Quantum Optics. Despite the field quantization in
1925, quantum optical effects were observed only seven decades after, the
first of them being the antibunching effect, as predicted by Carmichael and
Walls in 1976 \cite{1}, experimentally confirmed by Kimble, M. Dagenais and
L. Mandel \cite{2} in 1977. A second nonclassical effect was observed in
1985 by Slusher et al. \cite{3}, theoretically anticipated by Stoler et al. 
\cite{4} in 1970. A third one, the oscillations in the photon statistical
distribution, was observed in 1987 by Rempe et al. \cite{5}. Since then,
various nonclassical states of the quantized electromagnetic field were
studied, including their practical realization in laboratories in different
systems - one of them being the famous Schr\"{o}dinger ``cat" state, its
generation being sugested by Yurke and Stoler \cite{6}, Davidovich et al. 
\cite{7}, etc; its first experimental observation was obtained by the group
of Haroche \cite{8}. More recently, the community became aware of the first
experimental observation of the decoherence of the Schr\"{o}dinger ``cat" state,
in both realms of optical \cite{8.1} and atomic physics \cite{8.2}, and
constituting the first observation of the passage through the frontier that
separates the quantum and classical physics. After that, another interesting
topic emerged, as the quantum teleportation of states, first suggested by
Bernnett et al. \cite{10}, based on the nonlocal character of quantum
mechanics and contextualized by the EPR entangled states \cite{10.5}. This
somewhat ``bizarre" effect was first observed experimentally in 1997, by the
group of Zeilinger \cite{11}, concerning the teleportation of a single
photon state; the effect was later extended for atomic states and also for a
huge quantity of photons \cite{12}. Then, several publications in this line
appeared in the literature \cite{13,13.1,13.2,13.3}.

Besides the nonclassical effects of light field states, many researchers
became interested in the study of new states and new effects they could
exhibit, mainly concerning with their potential applications \cite{cv1}.
Then, it became also interesting the study of various schemes for the
generation of nonclassical light states \cite{114,14,cv2}. To this end, two
lines of study emerged: (i) when the issue is concerned with a state of a
stationary field, inside a high-Q microwave cavity; (ii) when concerning
with a traveling field, either thoughout the free space or a medium (optical
fibers, beam splitters, prisms, etc.). In both cases various proposals
appeared in the literature \cite{15a,15ab,15,15.1}. The extension of these
investigations for atomic systems has been also implemented. In this case
the system no longer concerns with traveling fields or a field trapped
inside a high-Q cavity; instead, it consists of atoms either inside or
crossing a cavity, including atoms inside a magneto-optical trap \cite%
{16,16.1}.

When focusing either the field or the atomic case the theoretical strategy
starts from a Hamiltonian describing the atom-field system, traditionally
treated via the Jaynes-Cummings model and the atom-field coupling usually
considered as a constant parameter. Comparatively, the number of such works
in the literature is very small when one considers the atom-field coupling
and/or atomic frequency as a time dependent parameter \cite{law,l1,l22,l2,l3}%
, including the case of time dependent amplitude \cite{abdalla}.
Nevertheless, this scenario is also relevant; \ for example, the state of
two qubits (qubits stand for quantum bits) with a desired degree of
entanglement can be generated via a time dependent atom-field coupling \cite%
{olaya}; such coupling can modify the dynamical properties of the atom and
the field, with transitions that involve a large number of photons \cite%
{yang}. In general, these studies are simplified by neglecting the atomic
decay from an excited level. Theoretical treatments taking into account this
complication of the real world also employs the Jaynes-Cummings model. In
these case, as expected, one finds decoherence of the state describing the
system, since the presence of dissipation destroys the state of a system as
time flows.

Here, taking advantage of what we have learned on the atom-field
interaction, we will study an advantageous system in practice (due to its
rapid response and better controllability \cite{y1}) by considering a
nanomechanical resonator (NR) interacting with a Cooper pair box (CPB). This
nanodevice has its own interest since its macroscopic nature and peculiar
effects of low-frequency noise in the solid-state impose obstacles requiring
more careful studies than a mere translation from quantum optics. It has
been explored in the study of quantum nondemolition measurements \cite%
{ak1,e1}, in the study of decoherence of nonclassical states, as Fock states
and superposition or entangled states describing mesoscopic systems \cite{e2}%
, etc. The fast advance in the tecnique of fabrication in nanotecnology
implied great interest in the study of the NR system in view of its
potential modern applications, as a sensor, largely used in various domains,
as in biology, astronomy, quantum computation, and more recently in quantum
information \cite{ak2} to implement the quantum qubit \cite{ak3} and in the
production of nonclassical states, e.g.: Fock states \cite{akk},
Schr\"{o}dinger's ``cat" states \cite{akk2}, squeezed states \cite{a34}, clusters
states \cite{ak}, etc. In particular, when accompanied by superconducting
charge qubits, the NR has been used to prepare entangled states \cite{akk1}.
Zhou et al.\cite{a34} have proposed a scheme to prepare squeezed states
using a NR coupled to a CPB qubit; in this proposal the NR-CPB coupling is
under an external control while the connection between these two interacting
subsystems play an important role in quantum computation. Such a control is
achieved via convenient change of system parameters, which can set ``on" and
``off" the interaction between the NR and the CPB, on demand.

One of the desired goals in this report is to verify the behavior and
properties of an entangled state describing the CPB-NR system, via the
Jaynes-Cummings model, by considering the energy dissipation in the CPB
during its transitions from an excited level to a ground state. Another
target is to verify if, and in which way, the time dependence of the CPB-NR
coupling modifies the dynamical properties of the state describing a
subsystem. We will also study the time evolution of the quantum entropy and
its power spectrum, as well as the CPB\ excitation inversion. There are some
evidences of entropy production, including the fact that the power spectrum
of stationary systems and subsystems can be used as dynamical criteria for
quantum caos \cite{Avi,avi1}. For the entropy power spectrum, such criteria
embody those already discussed in the literature concerned with fixed
parameters. Then, it seems adequate to look at the various characteristics
of the entropy to formulate a reasonable and suficient universal dynamical
criterium for the quantum caos. The degree of entanglement, represented by
the entropy in certain circumstances, has also shown itself being sensible
to the presence of a classical caos \cite{k2,k3}.

\section{Model hamiltonian for the CPB-NR system}

\begin{figure}[tbh]
\centering  
\fbox{\includegraphics[width=10cm, height=9cm]{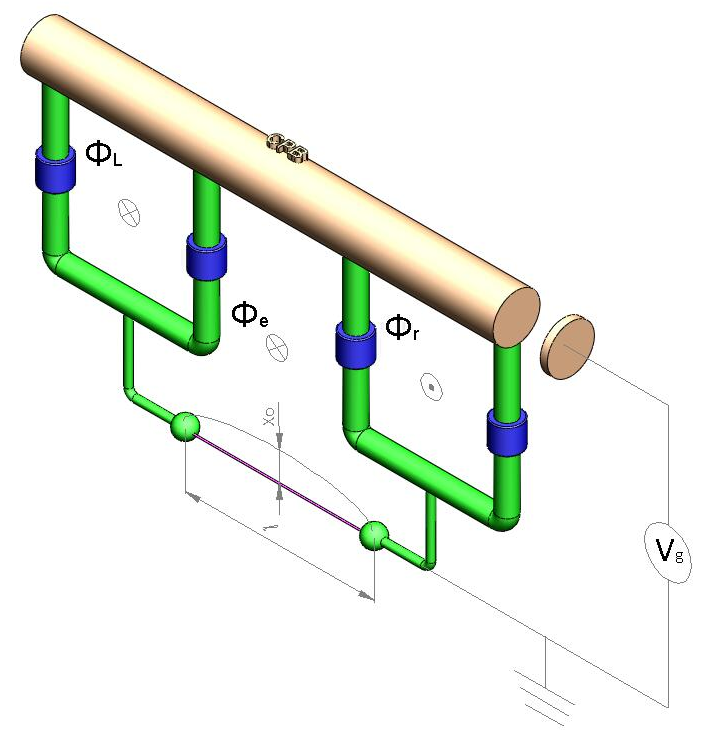}}
\caption{\textit{Model for the CPB-NMR coupling.}}
\label{cooper}
\end{figure}

There exist in the literature a large number of devices using the
SQUID-base, where the CPB charge qubit consists of two superconducting
Josephson junctions in a loop. In the present model a CPB is coupled to a NR
as shown in Fig. (\ref{cooper}); the scheme is inspired in the works by
Jie-Qiao Liao et al. \cite{ak3} and Zhou et al. \cite{a34} where we have
substituted each Josephson junction by two of them. This creates a new
configuration including a third loop. A superconducting CPB charge qubit is
adjusted via a voltage $V_{1}$\ at the system input and a capacitance $C_{1}$%
. We want the scheme ataining an efficient tunneling effect for the
Josephson energy. In Fig.(\ref{cooper}) we observe three loops: one great
loop between two small ones. This makes it easier controlling the external
parameters of the system since the control mechanism includes the input
voltage $V_{1}$ plus three external fluxes $\Phi (\ell ),$ $\Phi (r)$ and $%
\Phi _{e}(t)$. In this way one can induce small neighboring loops\emph{.}
The great loop contains the NR and its effective area in the center of the
apparatus changes as the NR oscillates, which creates an external flux $\Phi
_{e}(t)$ that provides the CPB-NR coupling to the system. %
%
In this work we will assume the four Josephson junctions being identical,
with the same Josephson energy $E_{J}^{0}$, the same being assumed for the
external fluxes $\Phi (\ell )$ and $\Phi (r)$, i.e., with same magnitude,
but opposite sign: $\Phi (\ell )=-\Phi (r)=\Phi (x)$. In this way, we can
write the Hamiltonian describing the entire system as

\begin{equation}
\hat{H}=\omega \hat{a}^{\dagger }\hat{a}+4E_{c}\left( N_{g}-\frac{1}{2}%
\right) \hat{\sigma}_{z}-4E_{J}^{0}\cos \left( \frac{\pi \Phi _{x}}{\Phi _{0}%
}\right) \cos \left( \frac{\pi \Phi _{e}}{\Phi _{0}}\right) \hat{\sigma}_{x},
\label{a1}
\end{equation}%
where $\hat{a}^{\dagger }(\hat{a})$ is the creation (annihilation) operator
for the excitation in the NR, corresponding with the frequency $\omega $ \
and mass $m$; $E_{J}^{0}$ and $E_{c}$ are respectively the energy of each
Josephson junction and the charge energy of a single electron; $C_{1}$ and $%
C_{J}^{0}$ stand for the input capacitance and the capacitance of each
Josephson tunel, respectively. $\Phi _{0}=h/2e$ is the quantum flux and $%
N_{1}=C_{1}V_{1}/2e$ is the charge number in the input with the input
voltage $V_{1}$. We have used the Pauli matrices to describe our system
operators, where the states $\left\vert g\right\rangle $ and $\left\vert
e\right\rangle $ (or 0 and 1) represent the number of extra Cooper pairs in
the superconduting island. We have: $\hat{\sigma}_{z}=\left\vert
g\right\rangle \left\langle g\right\vert -\left\vert e\right\rangle
\left\langle e\right\vert $, $\hat{\sigma}_{x}=\left\vert g\right\rangle
\left\langle e\right\vert -\left\vert e\right\rangle \left\langle
g\right\vert $ and $E_{C}=e^{2}/\left( C_{1}+4C_{J}^{0}\right) .$

The magnectic flux can be written as the sum of two terms, 
\begin{equation}
\Phi _{e}=\Phi _{b}+B\ell \hat{x}\text{ },  \label{a4}
\end{equation}%
where the first term $\Phi _{b}$ is the induced flux, corresponding to the
equilibrium position of the NR and the second term describes the
contribution due to the vibration of the NR; $B$ represents the magnectic
field created in the loop. We have assumed the displacement $\hat{x}$
described as $\hat{x}=x_{0}(\hat{a}^{\dagger }+\hat{a})$, where $x_{0}=\sqrt{%
m\omega /2}$ is the amplitude of the oscillation.

Substituting the Eq.(\ref{a4}) in Eq.(\ref{a1}) and controlling the flux $%
\Phi _{b}$ we can adjust $\cos \left( \frac{\pi \Phi _{b}}{\Phi _{0}}\right)
=0$ to obtain 
\begin{equation}
\hat{H}=\omega \hat{a}^{\dagger }\hat{a}+4E_{c}\left( N_{g}-\frac{1}{2}%
\right) \hat{\sigma}_{z}-4E_{J}^{0}\cos \left( \frac{\pi \Phi _{x}}{\Phi _{0}%
}\right) \sin \left( \frac{\pi B\ell \hat{x}}{\Phi _{0}}\right) \hat{\sigma}%
_{x}  \label{a8}
\end{equation}%
and making the approximation $\pi B\ell x/\Phi _{0}<<1$ we find 
\begin{equation}
\hat{H}=\omega \hat{a}^{\dagger }\hat{a}+\frac{1}{2}\omega _{0}\hat{\sigma}%
_{z}+\lambda _{0}(\hat{a}^{\dagger }+\hat{a})\hat{\sigma}_{x},  \label{a9}
\end{equation}%
where the constant coupling $\lambda _{0}=-4E_{J}^{0}\cos \left( \frac{\pi
\Phi _{x}}{\Phi _{0}}\right) \left( \frac{\pi B\ell x_{0}}{\Phi _{0}}\right) 
$ and the effective energy $\omega _{0}=8E_{c}\left( N_{g}-\frac{1}{2}%
\right) .$ In the rotating wave approximation the above Hamiltonian results
as 
\begin{equation}
\hat{H}=\omega \hat{a}^{\dagger }\hat{a}+\frac{1}{2}\omega _{0}\hat{\sigma}%
_{z}+\lambda _{0}(\hat{\sigma}_{+}\hat{a}+\hat{a}^{\dagger }\hat{\sigma}%
_{-}).
\end{equation}

Next, we will consider a more general scenario by substituting $\omega
\rightarrow \omega (t)=\omega +f\left( t\right) $ and $\lambda
_{0}\rightarrow \lambda (t)=\lambda _{0}\left[ 1+f\left( t\right) /\omega %
\right] $ \cite{yang,jf}; in addition, we assume the presence of a constant
decay rate $\gamma $ in the CPB, from its excited level to the ground state; 
$\omega _{0}$ is the transition frequency of the CPB and $\lambda _{0}$
stands for the CPB-NR\ coupling. $\hat{\sigma}_{\pm }$ and $\hat{\sigma}_{z}$
are the CPB transition and excitation inversion operators, respectively;
they act on the Hilbert space of atomic states and satisfy the commutation
relations $\left[ \hat{\sigma}_{+},\hat{\sigma}_{-}\right] =\hat{\sigma}_{z}$
and $\left[ \hat{\sigma}_{z},\hat{\sigma}_{\pm }\right] =\pm \hat{\sigma}%
_{\pm }$. As well known, the coupling parameter $\lambda (t)$ is
proportional to $\sqrt{\upsilon \left( t\right) /V\left( t\right) }$, where
the time dependent quantization volume $V\left( t\right) $ takes the form $%
V\left( t\right) =V_{0}/\left[ 1+f\left( t\right) /\omega \right] $ \cite%
{scully,l2,jf}. Accordingly, we obtain the new (\textit{non hermitean})
Hamiltonian

\begin{equation}
\hat{H}=\omega (t)\hat{a}^{\dagger }\hat{a}+\frac{1}{2}\omega _{0}\hat{\sigma%
}_{z}+\lambda (t)(\hat{\sigma}_{+}\hat{a}+\hat{a}^{\dagger }\hat{\sigma}%
_{-})-i\frac{\gamma }{2}\left\vert e\right\rangle \left\langle e\right\vert .
\label{b1}
\end{equation}

Non hermitean Hamiltonians (NHH) have been largely used in the literature.
As some few examples we mention: Ref. \cite{nh5}, where the authors use a
NHH and an algorithm to generalize the conventional theory; Ref. \cite{nh1},
using a NHH to get information about entrance and exit channels; Ref. \cite%
{nh6},\ using non hermitean techniques to study canonical transformations in
quantum mechanics; Ref. \cite{nh7}, solving quantum master equations in
terms of NHH; Ref. \cite{nh3}, using a new approach for NHH to study the
spectral density of weak H-bonds involving damping; Ref. \cite{nh8}, studing
NHH with real eighenvalues; Ref. \cite{nh4}, using a canonical formulation
to study dissipative mechanics exhibing complex eigenvalues;\ Ref. \cite{nh9}%
, studing NHH in non commutative space, and more recently: Ref. \cite{nh10},
studing the optical realization of relativistic NHH; Ref. \cite{l2}, studing
the evolution of entropy of atom-field interation; Ref. \cite{l22}, using a
damping JC-Model to study entanglement between two atoms, each one inside
distinct cavities

\section{Solving the CPB-NR system}

Now, the state describing our time dependent system can be written as

\begin{equation}
\left\vert \Psi \left( t\right) \right\rangle =\sum\nolimits_{n=0}^{\infty
}(C_{g,n}\left( t\right) \left\vert g,n\right\rangle +C_{e,n}\left( t\right)
\left\vert e,n\right\rangle ).  \label{b2}
\end{equation}%
Taking the CPB initially prepared in its excited state $\left\vert
e\right\rangle $ and the NR in a superposition of two coherent states, $%
\left\vert \beta \right\rangle =\eta (\left\vert \alpha \right\rangle
+\left\vert -\alpha \right\rangle )$, and expanding each coherent state
component in the Fock's basis, i.e., $\left\vert \alpha \right\rangle
=exp(-|\alpha |^{2}/2)\sum_{n=o}^{\infty }(\alpha ^{n}/\sqrt{n!})|n\rangle $%
,\ we have $\left\vert \beta \right\rangle =\sum\nolimits_{n=0}^{\infty
}F_{n}\left\vert n\right\rangle ,$where $\eta =[2+2\exp (-2\alpha
^{2})]^{-1/2}$ is the normalization factor.\ Assuming the NR and CPB
decoupled at $t=0$ and the initial conditions $C_{g,n}\left( 0\right) =0$
and $\sum\nolimits_{n=0}^{\infty }\left\vert C_{e,n}\left( 0\right)
\right\vert ^{2}=1$ we may write the Eq. (\ref{b2}) as

\begin{equation}
\left\vert \Psi \left( 0\right) \right\rangle =\sum\nolimits_{n=0}^{\infty
}F_{n}\left\vert e,n\right\rangle .  \label{b4}
\end{equation}

The time dependent Schr\"{o}dinger equation for the present system is

\begin{equation}
i\dfrac{d\left\vert \Psi \left( t\right) \right\rangle }{dt}=\hat{H}%
\left\vert \Psi \left( t\right) \right\rangle ,  \label{b5}
\end{equation}%
with the Hamiltonian $\hat{H}$ given in Eq. (\ref{b1}).\emph{\ }Substituting
Eq.(\ref{b1}) in Eq.(\ref{b5}) we get the (coupled) equations of motion for
the probabilitity amplitudes $C_{e,n}(t)$ and $C_{g,n+1}(t)$:%
\begin{eqnarray}
\frac{\partial C_{e,n}(t)}{\partial t} &=&-in\omega (t)C_{e,n}(t)-\frac{i}{2}%
\omega _{0}C_{e,n}(t)-i\lambda (t)\sqrt{n+1}C_{g,n+1}(t)-\frac{\gamma }{2}%
C_{e,n}(t),  \label{b8} \\
\frac{\partial C_{g,n+1}(t)}{\partial t} &=&-i(n+1)\omega (t)C_{g,n+1}(t)+%
\frac{i}{2}\omega _{0}C_{g,n+1}(t)-i\lambda (t)\sqrt{n+1}C_{e,n}(t).
\label{b9}
\end{eqnarray}

The solutions of the coefficientes $C_{e,n}(t)$, $C_{g,n+1}(t)$ furnish the
quantum dynamical properties of the system, including the CPB-NR
entanglement.

For the cases $f(t)=0$\ and $f(t)=$ $const,$ the Eq.(\ref{b8}) and Eq.(\ref%
{b9}) are exactly soluble. We find, analytically,%
\begin{eqnarray}
C_{g,n+1}(t) &=&(1/\zeta )[-2i\lambda e^{-1/4\delta t}\left( e^{1/4\zeta
t}-e^{-1/4\zeta t}\right) \sqrt{n+1}F_{n}]{,}  \label{bb1} \\
C_{e,n}(t) &=&\left( 1/2\zeta \right) [ie^{-1/4\delta t}\left( e^{1/4\zeta
t}\left( i\gamma +2\omega -i\zeta -2\omega _{0}\right) -e^{-1/4\zeta
t}\left( i\gamma +2\omega +i\zeta -2\omega _{0}\right) \right) F_{n}]{,}
\label{bb2}
\end{eqnarray}%
where $\delta =\gamma +2i\omega (1+2n)$ and $\zeta =[\gamma (\gamma
+4i(\omega _{0}-\omega ))-4({\omega }^{2}+\omega _{0}^{2})-16{\lambda }%
^{2}(1+n)+8\omega \omega _{0}]^{1/2}.$ However, when the coupling $f(t)$ is
time dependent the solution of this system of equations is found only
numerically.

As well known, in the presence of decay rate $\gamma $ in the CPB the state
of the whole CPB-NR system becomes mixed. In this case its description
requires the use of the density operator $\hat{\rho}_{CN}$, which describes
the entire system. To obtain the reduced \ density matrix describing the CPB
(NR) sub-system we must trace over variables of the NR (CPB) sub-system. For
example, 
\begin{equation}
\hat{\rho}_{NR}=Tr_{CPB}(\hat{\rho}_{CN})=\sum\limits_{n}\sum\limits_{n^{%
\prime }}\left[ C_{e,n}(t)C_{e,n^{\prime }}^{\ast
}(t)+C_{g,n}(t)C_{g,n^{\prime }}^{\ast }(t)\right] \left\vert n\right\rangle
\left\langle n\prime \right\vert .  \label{c2}
\end{equation}

\section{Entropy of sub-systems}

Recently, researchers have employed several methods to study the dynamical
of entanglement \cite{l22,l2,l3,k41,k6}. As proved by Phoenix and Knight 
\cite{k4} the von Neumann entropy offers a quantitative measure of disorder
of a system and of the purity of a quantum state. Such entropy, defined as $%
S_{NR(C)}=-Tr(\hat{\rho}_{N(C)}\ln \hat{\rho}_{N(C)})$, is a measure that is
sensible to quantum entanglement of two interacting subsystems. The quantum
dynamics described by the Eq. (\ref{b1}) furnishes the CPB-NR entanglement
and we will employ the von Neumann quantum entropy as a measure of the
degree of entanglement. The entropy $S$ of a quantum system, when composed
of two subsystems, obeys a theorem due to Araki and Lieb, which stablishes
that: $\left\vert S_{CPB}-S_{NR}\right\vert \leq S\leq S_{CPB}+S_{NR}$; $%
S_{CPB}$ and $S_{NR}$ standing for the entropies of the subsystems. $S$
stands for the total entropy of CPB-NR system. One immediate consequence of
the above inequality is that, if one prepares the entire system in a pure
state at $t=0$, then both components of the whole system have the same
entropy for the subsequent time evolution.\emph{\ }So, when assuming our
system initially in a pure and decoupled state the entropies of the CPB and
NR become identical, namely, $S_{CPB}(t)=S_{NR}(t).$ Then, one only needs to
calculate the quantum entropy of a subsystem to get its entanglement
evolution. We obtain, from the Eqs. (\ref{c2}) and $S_{NR}=-Tr(\hat{\rho}%
_{NR}\ln \hat{\rho}_{NR})$,

\begin{equation}
S_{NR}(t)=-\left[ \wedge _{NR}^{+}(t)\ln (\wedge _{NR}^{+}(t))+\wedge
_{NR}^{-}(t)\ln (\wedge _{NR}^{-}(t))\right] ,  \label{c3}
\end{equation}%
where,

\begin{equation}
\wedge _{NR}^{\pm }(t)=\frac{1}{2}\left( 1\pm \sqrt{(\left\langle
R_{1}|R_{1}\right\rangle -\left\langle R_{2}|R_{2}\right\rangle
)^{2}+4\left\vert \left\langle R_{1}|R_{2}\right\rangle \right\vert ^{2}}%
\right) ,  \label{c4}
\end{equation}%
with \ $\left\langle R_{1}|R_{1}\right\rangle =\sum_{n=0}^{\infty
}\left\vert C_{e,n}(t)\right\vert ^{2},~$\ $\left\langle
R_{2}|R_{2}\right\rangle =\sum_{n=0}^{\infty }\left\vert
C_{g,n+1}(t)\right\vert ^{2}$ and$\ \left\langle R_{1}|R_{2}\right\rangle
=\left\langle R_{2}|R_{1}\right\rangle ^{\ast }=\sum_{n=0}^{\infty
}C_{e,n+1}^{\ast }(t)C_{g,n+1}(t).$

\begin{figure}[tbh]
\centering  
\fbox{\includegraphics[width=7cm, height=6cm]{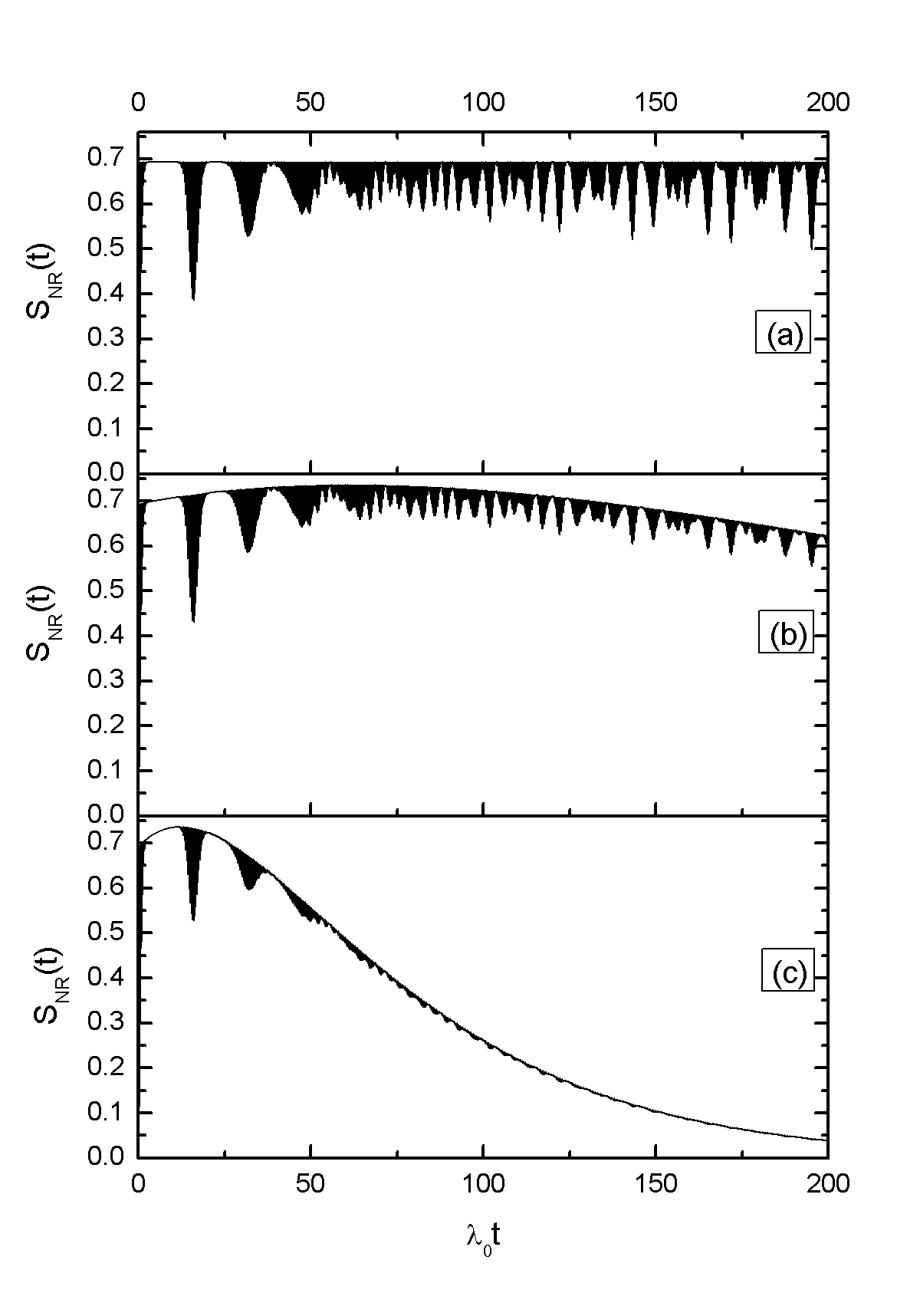}}
\caption{\textit{Time evolution of the Entropy when the NR is initially
prepared\emph{\ in an even ``cat-state}", for different values of the decay
rate $\protect\gamma $: (a) $\protect\gamma =0.0\protect\lambda _{0}$, (b) $%
\protect\gamma =0.01\protect\lambda _{0}$, and (c) $\protect\gamma =0.05%
\protect\lambda _{0}$, \ with $\protect\alpha =5$, $\protect\omega =\protect%
\omega _{0}=2000\protect\lambda _{0}$, $f(t)=0$.}}
\label{s4}
\end{figure}

\begin{figure}[tbh!]
\begin{minipage}[b]{0.48 \linewidth}
\fbox{\includegraphics[width=7cm, height=6cm]{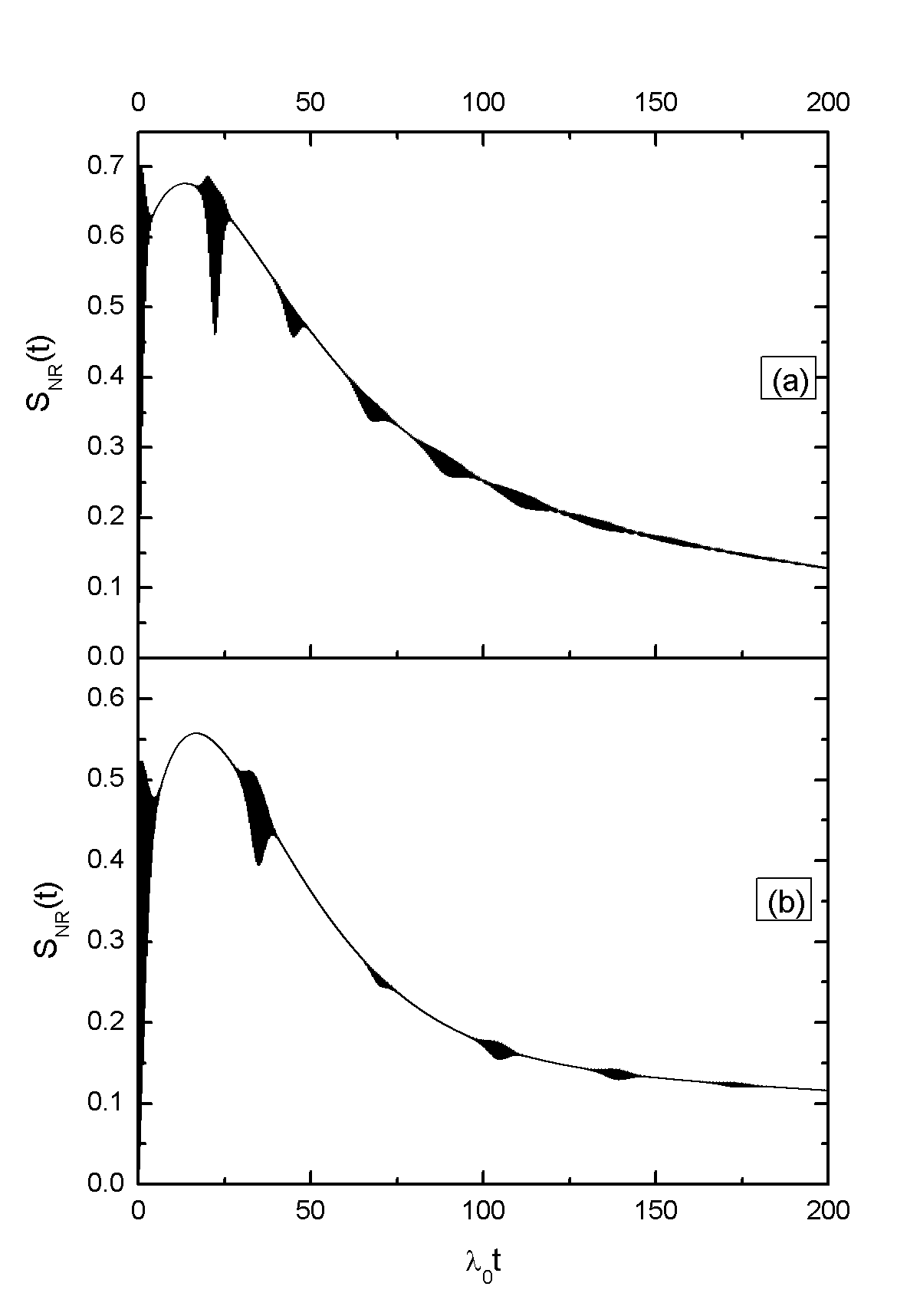}}\\
\caption{\textit{Same as in Fig.\protect\ref{s4}, \ now for different values
of detunings (cf. $f(t)=\Delta =const$): (a) $\Delta =10\protect\lambda %
_{0}$ and (b) $\Delta =20\protect\lambda _{0}$ , with $\protect\alpha =5$, $%
\protect\omega =\protect\omega _{0}=2000\protect\lambda _{0}$, $\protect%
\gamma =0.05\protect\lambda _{0}$.}}
\label{s5}
       \end{minipage}\hfill 
\begin{minipage}[b]{0.48 \linewidth}
\fbox{\includegraphics[width=7cm, height=6cm]{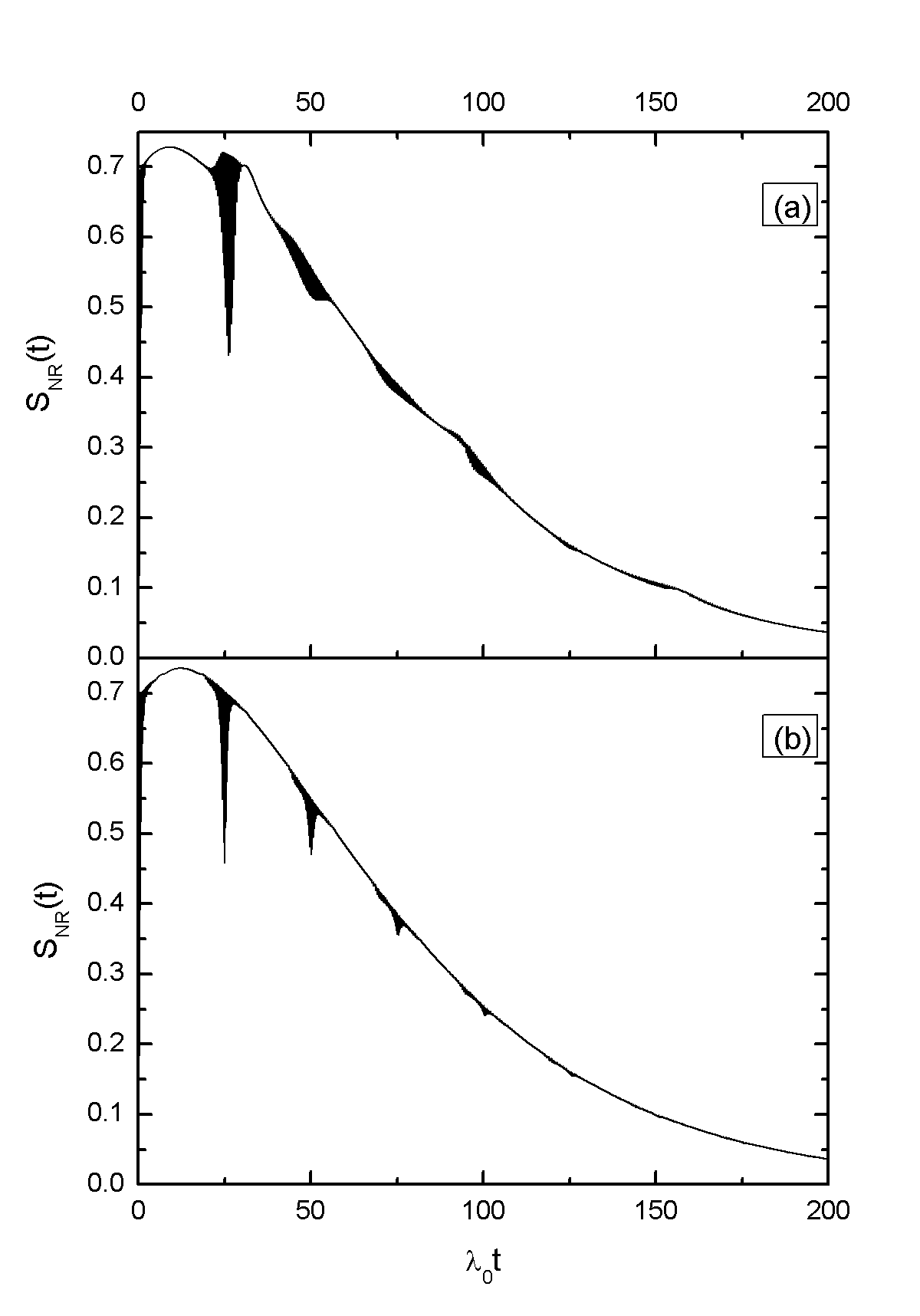}}\\
\caption{\textit{\ Same as in Figs. \protect\ref{s4} and \protect\ref{s5},
for time-dependent detunings (cf. $f(t)=c\sin (\protect\omega \prime t)$ ):
with (a) $c=20\protect\lambda _{0}$ e $\protect\omega \prime =0.1\protect%
\lambda _{0}$, and (b) $c=20\protect\lambda _{0}$ e $\protect\omega \prime
=0.5\protect\lambda _{0}$}}
\label{s6}
       \end{minipage}
\end{figure}

We can now look at the time evolution of the NR entropy. We will assume the
NR subsystem initially in an even ``Schr\"{o}dinger-cat" state. Firstly we
consider the resonant case $(f(t)=0);$ the time evolution of the NR entropy
with different decay rates $\gamma $ in the CPB\textbf{, }with\textbf{\ }$%
\omega =\omega _{0}=2000\lambda _{0}$ and the ``cat"-state with $\alpha =5$,
as shown in Figs. \ref{s4}(a), \ref{s4}(b), and \ref{s4}(c). In an ideal
case the CPB decay rate vanishes. As displayed in Fig. \ref{s4}(a) the
maximum value of the entropy of the NR is close to $\ln 2$. Just after the
start of the CPB-NR interaction the entropy of the NR stabilizes at this
value by a small interval and then recovers the oscillations as time goes
on. The CPB sub-system is stable while standing in its ground state $%
\left\vert g\right\rangle $; but when lying in its excited state $\left\vert
e\right\rangle ,$ various factors as spontaneous emission among others,
imply its decay to the ground state. For a small decay rate (see Fig. \ref%
{s4}(b) the maximum entanglement becomes significant only for large times.
However, the increase of the decay rate produces a drastic change on the
entanglement (see Fig. \ref{s4}(c), with a great reduction in their swings,
leading the CPB to its ground state (zero entropy). This effect upon the
entropy of the CPB also affects the entropy of the NR (Figs. \ref{s4}). %
%
Secondly, we modify the previous case by including the presence of a
detuning $(f(t)=\Delta \neq 0)$ to verify its influence opon our interacting
system. We take the decay rate as $\gamma =0.05\lambda _{0}$, with $%
f(t)=\Delta =const$ and $\Delta \ll \omega _{0}$, $\omega $. As result the
entanglement remains for long time as the value of $\Delta $\ increases\emph{%
,} as we see comparing Fig. \ref{s5}(a) with Fig. \ref{s4}(c); this\ event
is accompanied by a diminution of the maximum entropy (see Figs \ref{s5}(a)
and \ref{s5}(b)). When the detuning increases, the CPB transitions decreases
(cf. Figs. \ref{inv5}).


Thirdly, we extend the detuning to the time dependent case, assuming $%
f(t)=c\sin (\omega \prime t)$, where $c$ and $\omega \prime $ are parameters
of amplitude and frequency modulation of the NR with the condition $\omega
\prime <c\ll \omega _{0}$\emph{, }$\omega $. Comparing the Fig. \ref{s6}(a)
with Fig. \ref{s5}(b) we see that the sinusoidal modulation does not favor
the entanglement for long time. However, the frequency modulation turns the
maxima of entanglement greater (see Fig. \ref{s6}). Comparing Fig. \ref{s6}%
(a) with \ref{s6}(b) we see that when the frequency $\omega \prime $ grows
the oscillations of the entropy decrease.

\section{Power Spectrum of the Entropy}

To get a better understanding of the entropy we have considered its power
spectrum (\textbf{PS}). It consists of a frequency-dependent function, being
real, positive, and constructed from the following Fourier transform \cite%
{scully}%
\begin{equation}
PS(\varpi )=\frac{1}{\pi }\int_{0}^{\tau _{\max }}S_{NR}(t)\exp (i\varpi
t)dt,  \label{c5}
\end{equation}%
where $\tau _{\max }=$ $\lambda _{0}t_{\max }$ stands for the maximum
interaction interval in the plot $S_{NR}(t)$ versus $\lambda _{0}t$.

\begin{figure}[tbh]
\centering  
\fbox{\includegraphics[width=7cm, height=6cm]{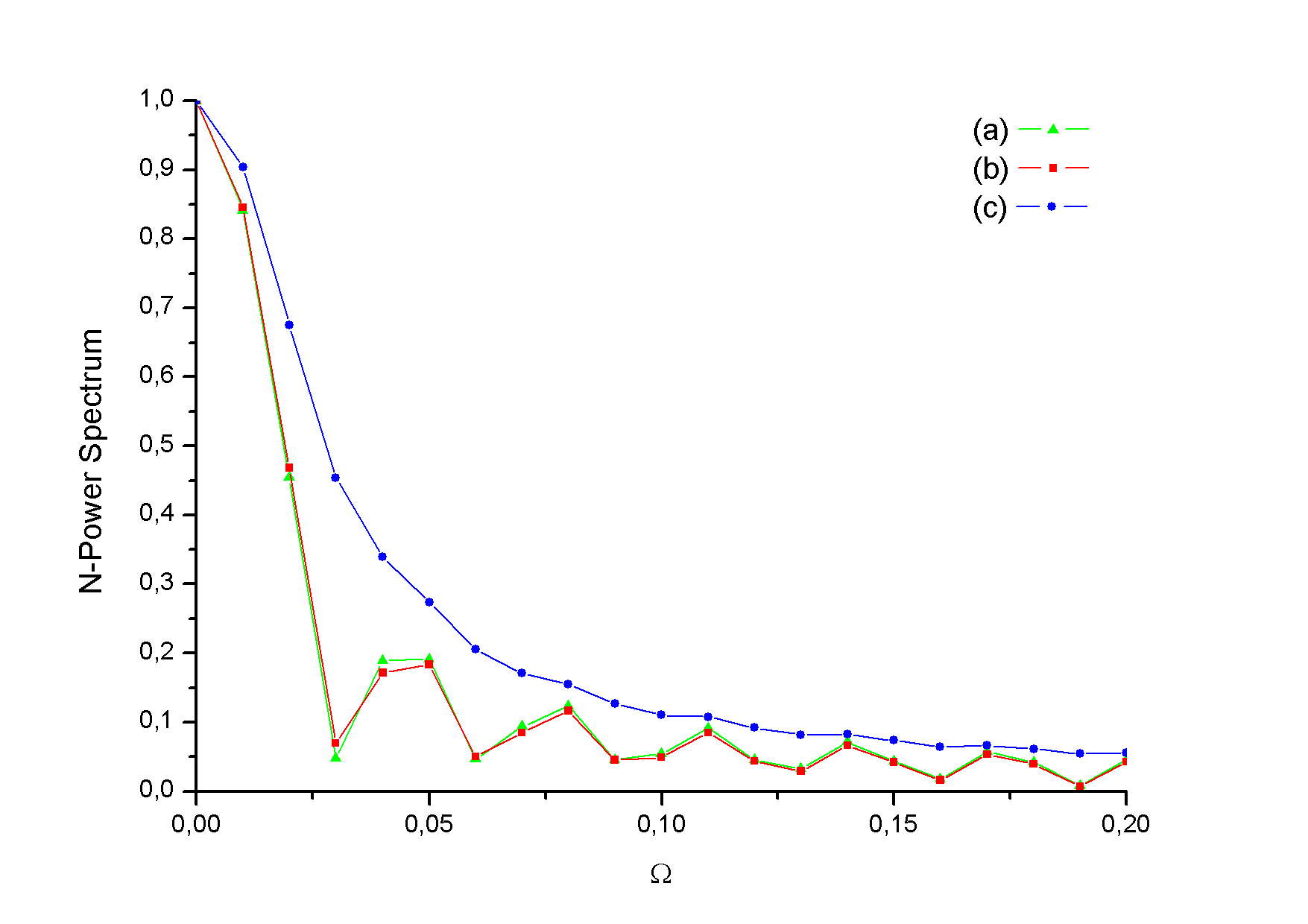}}
\caption{\textit{Time evolution of the Normalized Power Spectrum when the NR
is initially prepared in ``cat-state", concerning with the Entropies shown in
the plots of Fig. \protect\ref{s4}: the plot \protect\ref{ps4}(i) refers to
the plot \protect\ref{s4}(i), i = a, b, c.}}
\label{ps4}
\end{figure}

\begin{figure}[tbh!]
\begin{minipage}[b]{0.48 \linewidth}
\fbox{\includegraphics[width=7cm, height=6cm]{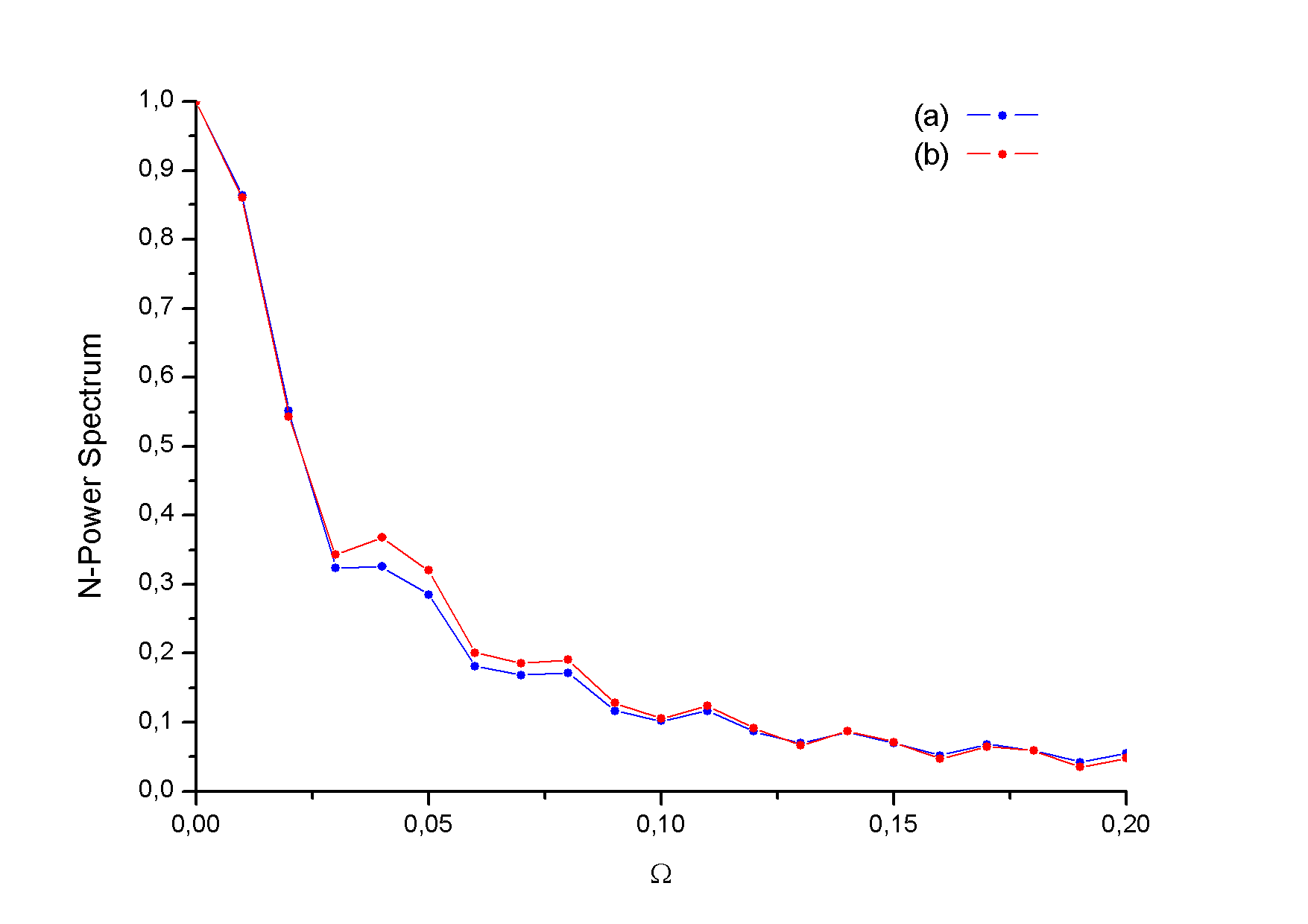}}\\
\caption{\textit{Same as in Fig. \protect\ref{ps4}, concerning with the
Entropies shown in the plots of Fig. \protect\ref{s5}: the plot \protect\ref%
{ps5}(i) refers to the plot \protect\ref{s5}(i), i = a, b.}}
\label{ps5}
       \end{minipage}\hfill 
\begin{minipage}[b]{0.48 \linewidth}
\fbox{\includegraphics[width=7cm, height=6cm]{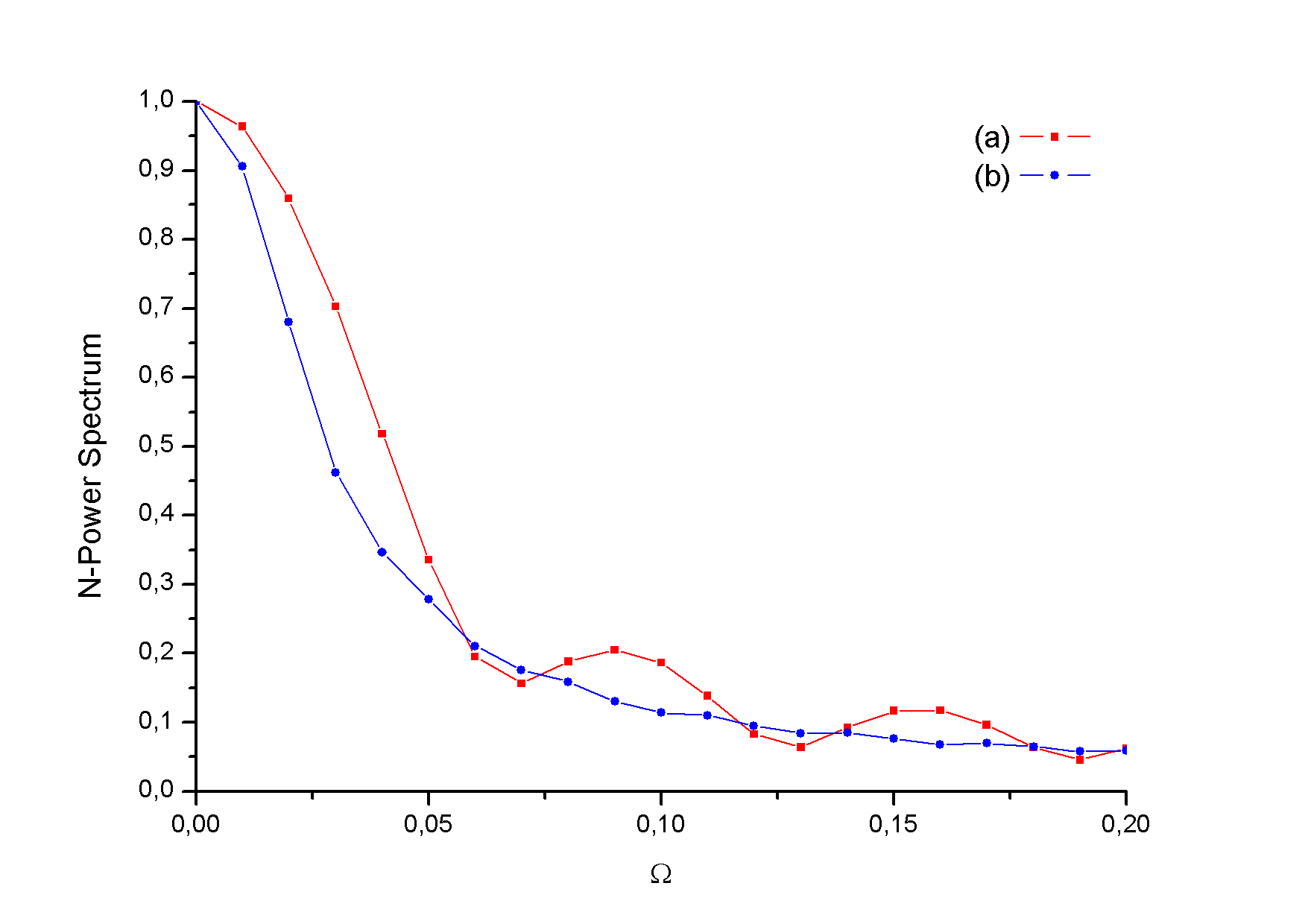}}\\
\caption{\textit{Same as in Figs. \protect\ref{ps4} and \protect\ref{ps5},
concerning with the Entropies shown in the plots of Fig. \protect\ref{s6}:
the plot \protect\ref{ps6}(i) refers to the plot \protect\ref{s6}(i), i = a,
b. }}
\label{ps6}
       \end{minipage}
\end{figure}

The entropy PS is obtained from the above equation and plotted in Figs. \ref%
{s4}(a), \ref{s4}(b) and \ref{s4}(c). As expected, the amplitude of
oscillations of this PS is reduced in the presence of growing decay rates
(see Figs. \ref{ps4}(a), \ref{ps4}(b) and \ref{ps4}(c) ); when we add a
constant detuning $($ $f(t)=\Delta \neq 0)$ and a decay rate $\gamma
=0.05\lambda _{0}$ we see in Figs. \ref{ps5}(a) and \ref{ps5}(b) the maximum
value of the PS increasing when $\Delta $ also increases. In the case $%
f(t)=csin(\omega \prime t)$ the frequency of entropy PS is smoothly
attenuated, with a peak around $\Omega =0.1,$ as shown in Fig. \ref{ps6}(a).
When the frequency $\omega \prime $ increases the entropy PS is rapidly
attenuated (cf. Figs. \ref{ps6}(a) and \ref{ps6}(b)). 


\section{Excitation Inversion of the CPB}

The CPB excitation inversion, $I_{CPB}(t)$, is an important observable of two
level systems. It is defined as the difference of the probabilities of
finding this system in the excited and in the ground state; for the CPB it
reads,%
\begin{equation}
I_{CPB}(t)=\sum\limits_{n=0}^{\infty }\left[ \left\vert
C_{e,n}(t)\right\vert ^{2}-\left\vert C_{g,n+1}(t)\right\vert ^{2}\right] .
\label{joao}
\end{equation}

\begin{figure}[tbh!]
\centering  
\fbox{\includegraphics[width=7cm, height=6cm]{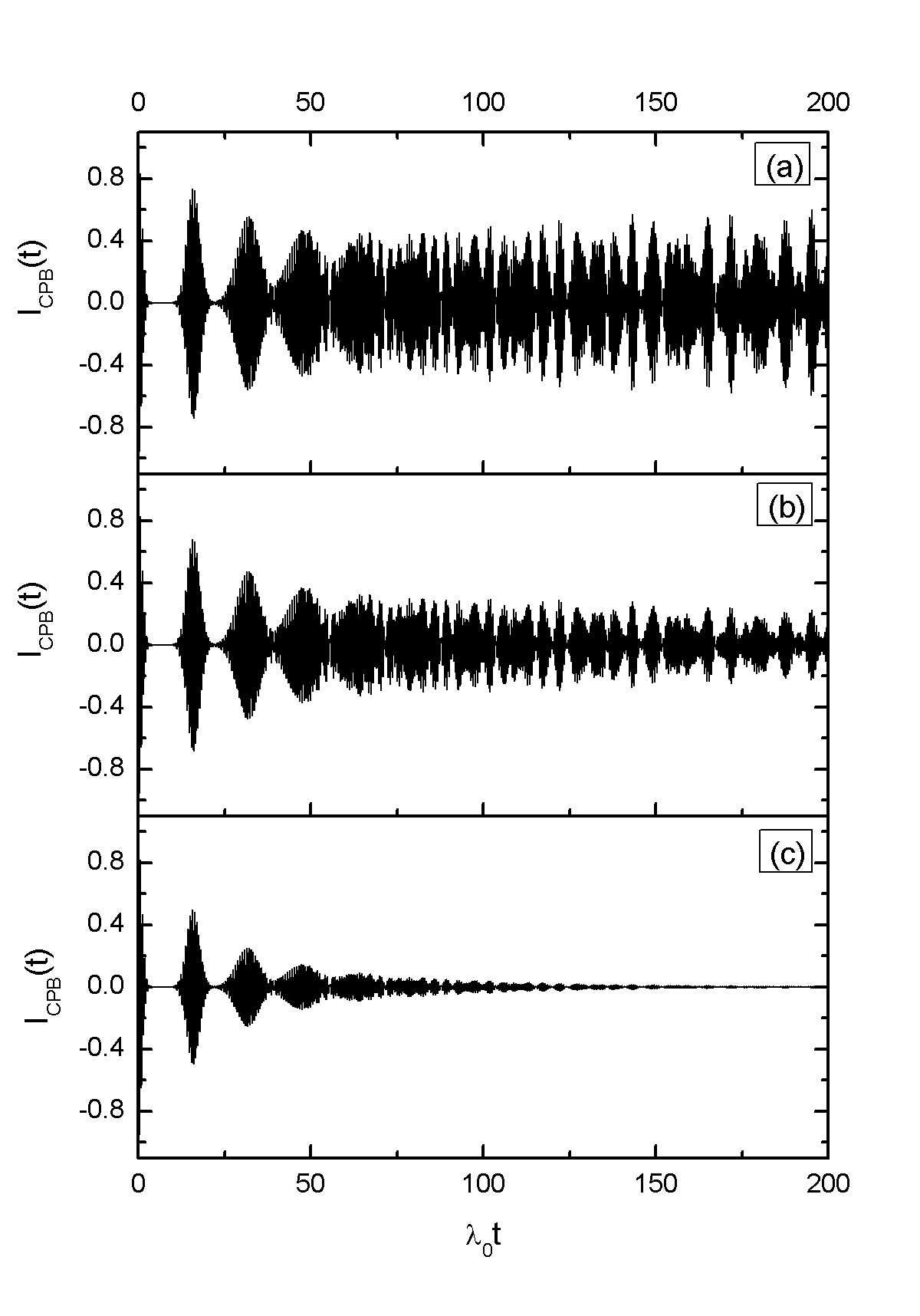}}
\caption{\textit{Time evolution of the CPB\ Excitation Inversion with the NR
initially prepared in the even ``cat-state", for various values of the decay
rate: (a) $\protect\gamma =0.0\protect\lambda _{0}$ (b) $\protect\gamma =0.01%
\protect\lambda _{0}$, and (c) $\protect\gamma =0.05\protect\lambda _{0}$,
with $\protect\alpha =5$, $\protect\omega =\protect\omega _{0}=2000\protect%
\lambda _{0}$, and $f(t)=0$ (resonance).}}
\label{inv4}
\end{figure}

\begin{figure}[tbh!]
\begin{minipage}[b]{0.48 \linewidth}
\fbox{\includegraphics[width=7cm, height=6cm]{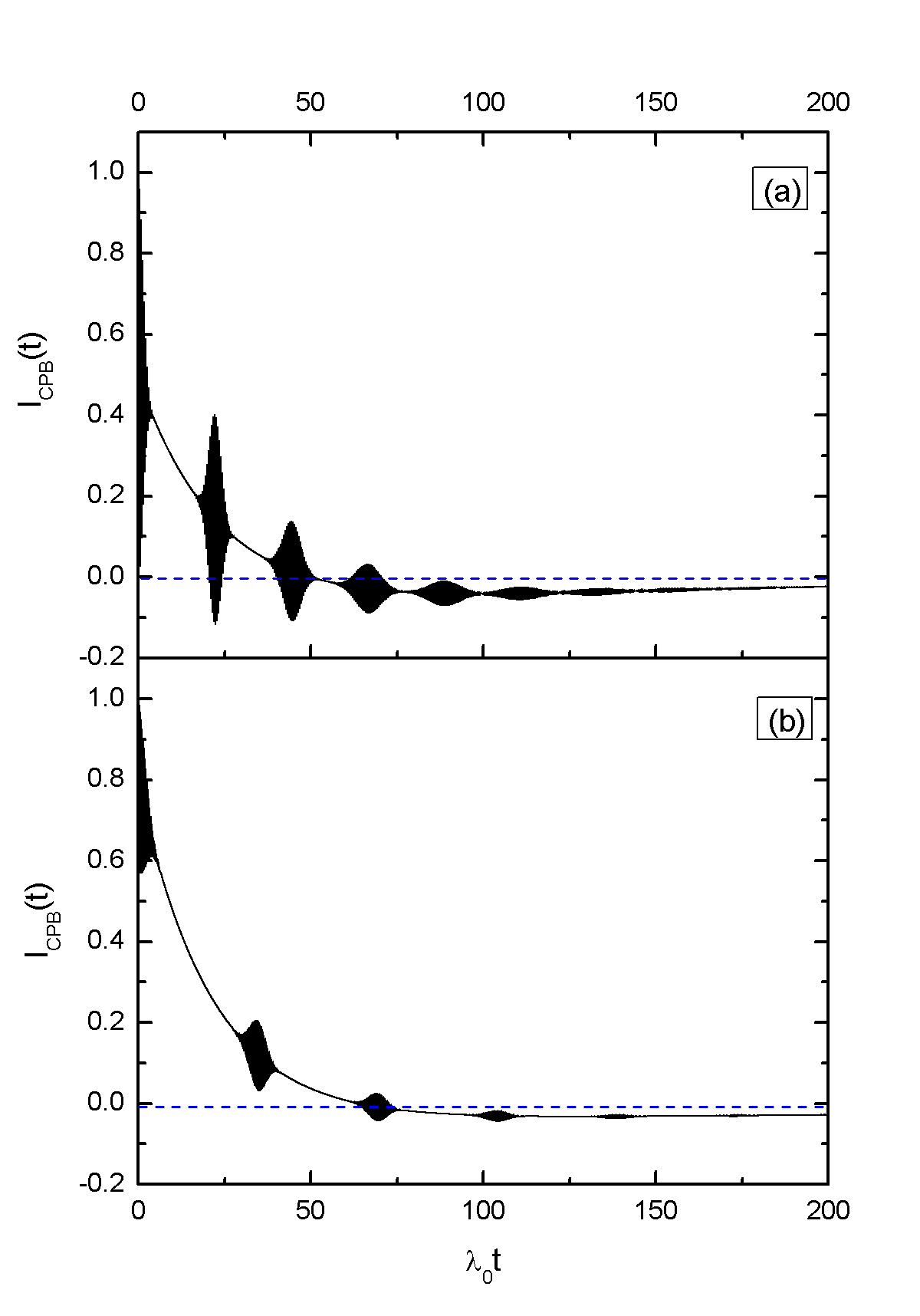}}\\
\caption{\textit{Same as in Fig. \protect\ref{inv4}, for diferent values of
detuning (cf. $f(t)=\Delta =const$): (a) $\Delta =10\protect\lambda _{0}$\
and (b) $\Delta =20\protect\lambda _{0}$, with $\protect\alpha =5$, $\protect%
\omega =\protect\omega _{0}=2000\protect\lambda _{0}$, $\protect\gamma =0.05%
\protect\lambda _{0}$.}}
\label{inv5}
       \end{minipage}\hfill 
\begin{minipage}[b]{0.48 \linewidth}
\fbox{\includegraphics[width=7cm, height=6cm]{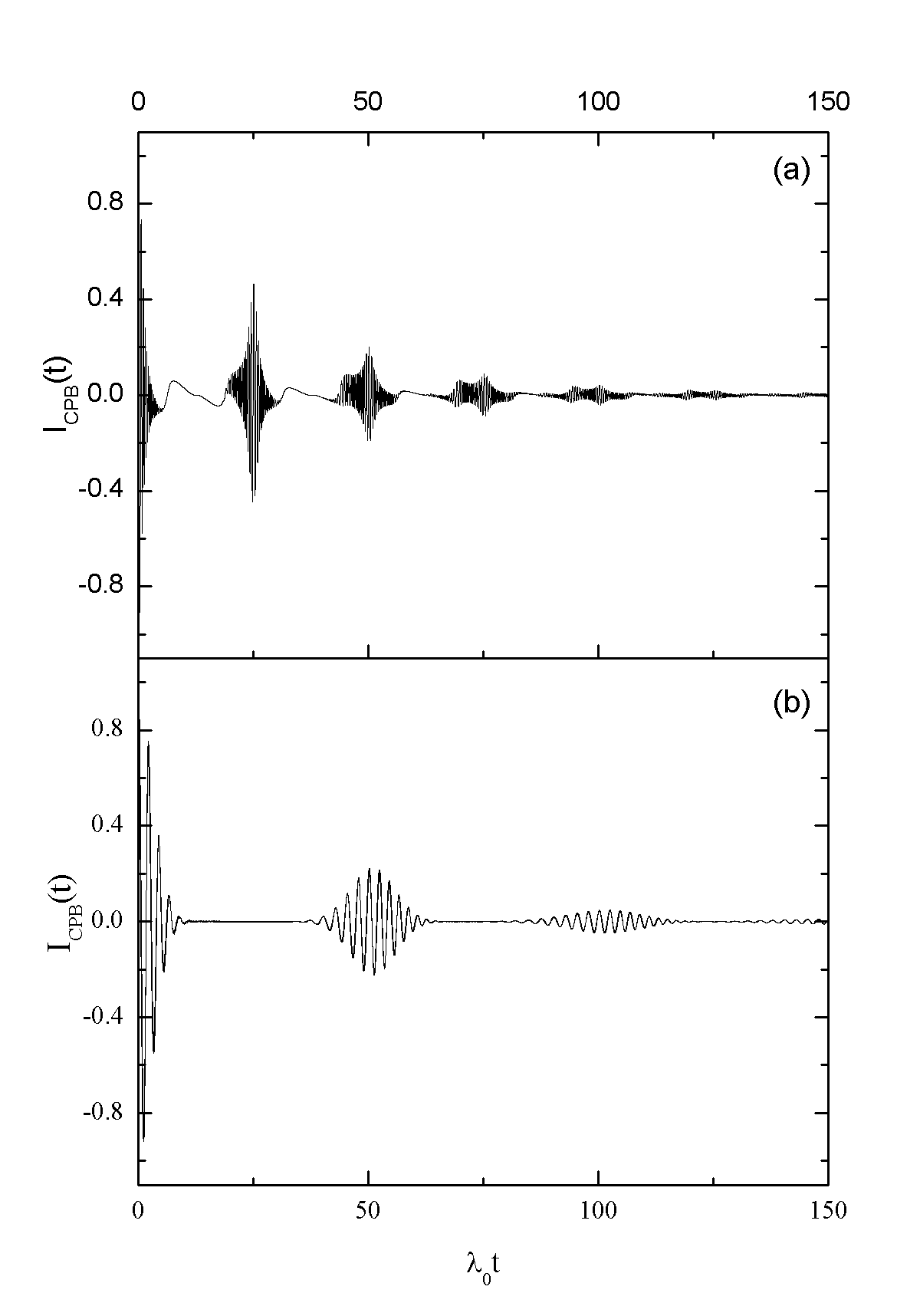}}\\
\caption{\textit{Same as in Figs. \protect\ref{inv4} and \protect\ref{inv5},
for time-dependent detunings (cf. $f(t)=c\sin (\protect\omega \prime t)$ ):
\ (a) $c=20\protect\lambda _{0}$, $\protect\omega \prime =0.5%
\protect\lambda _{0}$, (b) $c=60\protect\lambda _{0}$, $\protect\omega \prime =20%
\protect\lambda _{0}$, with $\protect\alpha =5$, $\protect\omega =\protect%
\omega _{0}=2000\protect\lambda _{0}$, $\protect\gamma =0.05\protect\lambda %
_{0}$.}}
\label{inv6}
       \end{minipage}
\end{figure}

The Eq. (\ref{joao}) allows us to look at the time evolution of the CPB
excitation inversion. First, we assume the resonant case $(f(t)=0),$ for
different values of the decay rate $\gamma $, with $\alpha =5$ and $\omega
=\omega _{0}=2000\lambda _{0}$ as in Fig. \ref{inv4}. Figs. \ref{inv4}(a),
(b) and (c) exhibit identical collapse and revival, but with different
amplitudes: the higher the decay rate, the lower the amplitude of
oscillations of CPB excitation inversion. However, in the presence of a
fixed detuning, with $f(t)=\Delta =const$ and $\Delta \ll \omega _{0},$ $%
\omega $, we see that CPB excitation inversion in Fig. \ref{inv5}(a) occurs
only inside the interval $\tau =\lambda _{0}t\in $ $(15,30)$; differently,
in Fig. \ref{inv5}(b) this event occurs inside the interval $\tau \in
(50,75) $, with amplitude smaller than that in Fig. \ref{inv5}(a); this is
the effect caused by a constant detuning upon the CPB excitation inversion.

For a time dependent detuning, $f(t)=csin(\omega \prime t)$, the frequency
of \ the CPB excitation inversion accompanies the frequency $\omega \prime $%
, as shown in Figs. \ref{inv6}. When we compare the Fig. \ref{inv6}(b) with
Fig. \ref{inv4}(c), we see: if $\omega \prime $\ increases, the interval
of collapse of the excitation inversion also increases. However, looking at
Figs. \ref{inv6}(a), \ref{inv6}(b) we see that the increasing of $\omega \prime $
as in Fig. \ref{inv6}(b) introduces equally spaced collapse intervals,
accompanied by revivals modulated by the parameter $\omega \prime $\ . Now, we
compare the CPB excitation inversion with constant detuning $(f(t)=\Delta
=const)$\ and with a time dependent detuning ($f(t)=csin(\omega \prime t)$):
looking at Figs. \ref{inv5} we see the plots of excitation inversion
showing neither collapses nor revivals, with exceptions of small regions
exhibinting excitation inversion (Fig.\ref{inv5}(a)); in Fig \ref{inv5}(b)
only a single such region appears. However, when considering a time dependent
detuning (Fig. \ref{inv6}(b)), it nicely restitutes those collapses and
revivals that appear in the resonant case (cf. Fig. \ref{inv4}(c)).


\section{Conclusion}

We have considered a Hamiltonian model for a CPB-NR interacting system to
study Entropy, its Power Spectrum and the CPB Excitation Inversion. These
properties characterize the entangled state that describes this coupled
system for various values of the parameters involved. We have included
dissipation and assumed the NR initially in a Schr\"{o}dinger ``cat"-state
and the CPB\ in excited state. We have also considered the following
scenarios: (\textit{i}) both subsystems in resonance (detuning $f=0$); (%
\textit{ii}) off-resonance, with a constant detuning $(f=\Delta \neq 0)$,
and (\textit{iii}) with a time dependent detuning $(f(t)=csin(\omega \prime
t))$. The results were discussed in the previous section. Concerning with
the entropy we see that when the NR is initially in a Schr\"{o}dinger
``cat"-state, the entropy lasts longer than in an atom-field system, with the
field initially in a coherent state (cf. Ref. \cite{l2}). Concerning the
Excitation Inversion, an interesting result emerges: although the presence
of a constant detuning destroys the collapse and revivals of the excitation
inversion, these effects are restituted by the action of convenient time
dependent detunings - even in the presence of damping. It is also worth
emphasizing that the presence of an external force upon the NR changes the
magnetic flux $\Phi _{e}$ (cf. Fig. \ref{cooper}), which provides the
control of the parameters $\omega (t)$ and $\lambda (t)$.

\section*{Acknowledgments}

The authors thank the FAPEG and CNPq, Brasilian Agencies, for partially
supporting this paper.


\end{document}